\documentclass[12pt,letterpaper]{article}
\usepackage[dvips]{graphicx}

\def\title#1{{\bf\Large #1}}
\def\sec#1{\vspace{1.00\baselineskip} \noindent {\bf\large #1}
\vspace{0.25\baselineskip}}

\def\d{\dagger}
\def\<{\langle}
\def\>{\rangle}

\def\tr{{\rm Tr}}

\def\be{\begin{equation}}
\def\ee{\end{equation}}
\def\ba{\begin{eqnarray}}
\def\ea{\end{eqnarray}}

\begin{document}


\begin{center}

\vspace*{2.00\baselineskip}

\title{Stokes-space formalism for Bragg scattering in a fiber} \\

\vspace{1.00\baselineskip}

C. J. McKinstrie \\
{\it\small Bell Laboratories, Alcatel--Lucent, Holmdel, New Jersey
07733}

\vspace{0.50\baselineskip}

Abstract \\

\parbox[t]{5.5in}{\small Optical frequency conversion by four-wave mixing
(Bragg scattering) in a fiber is considered. The evolution of this
process can be modeled using the signal and idler amplitudes, which
are complex, or Stokes-like parameters, which are real. The
Stokes-space formalism allows one to visualize power and phase
information simultaneously, and produces a simple evolution equation
for the Stokes parameters.}

\end{center}

\newpage

\sec{1. Introduction}

Parametric devices based on four-wave mixing (FWM) in fibers can
amplify, frequency convert, phase conjugate, regenerate and sample
optical signals in communication systems \cite{han02,mck07}. The
subject of this report is the nondegenerate FWM process called Bragg
scattering (BS), in which a sideband (signal) photon and a pump
photon are destroyed, and different sideband (idler) and pump
photons are created ($\pi_s + \pi_q \rightarrow \pi_i + \pi_p$,
where $\pi_j$ represents a photon with frequency $\omega_j$). The
frequencies of the interacting waves are illustrated in Fig. 1.
There is considerable interest in BS [3--14], because of its ability
to generate an idler whose frequency is tunable \cite{ino94}, and
which is not polluted by excess noise \cite{gna06}.
\begin{figure}[h]
\centerline{\includegraphics[width=2.4in]{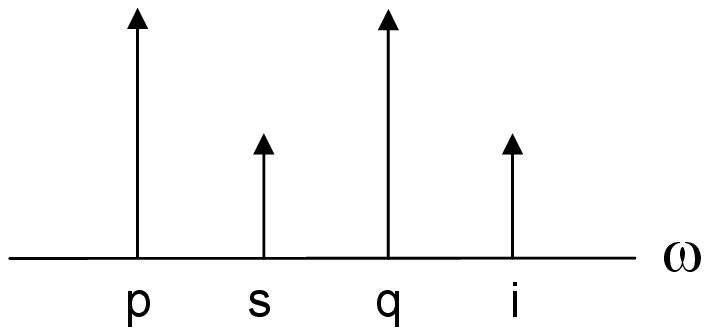}
\hspace{0.2in}
\includegraphics[width=2.4in]{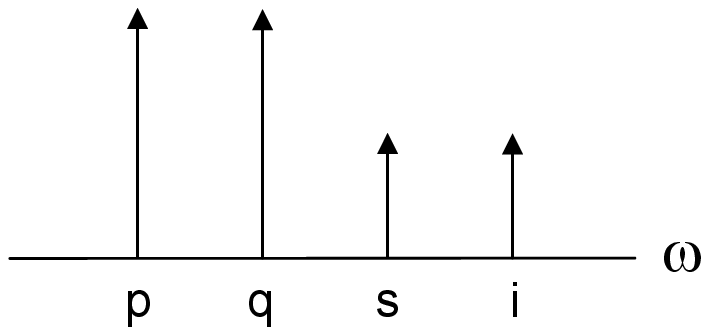}} \caption{Frequency
diagrams for Bragg scattering in a fiber. The long arrows denote
strong pumps ($p$ and $q$), whereas the short arrows denote weak
sidebands ($s$ and $i$).}
\end{figure}

BS is described by the sideband amplitudes $A_s$ and $A_i$. These
complex amplitudes are the components of a Jones vector and their
spatial evolution is a linear transformation in Jones space. Two
complex variables are equivalent to four real variables. However, an
inspection of the BS equations shows that one variable is ignorable,
so BS can be described by only three real variables, which are the
components of a Stokes-like vector. The Stokes-space description of
BS is useful because it allows one to visualize power and phase
information simultaneously, and because evolution in Stokes space is
a simple rotation of the Stokes vector.

This report is organized as follows: In Sec. 2 the coupled equations
for the sideband (mode) amplitudes are stated and solved. In Sec. 3
these scalar equations are rewritten as a matrix equation for the
Jones vector and their solutions are used to specify the associated
transfer matrix. In Sec. 4 the Stokes parameters associated with the
sideband amplitudes are defined and the (unknown) elements of a
Stokes-space rotation matrix are related to the (known) elements of
a Jones-space transfer matrix. An evolution equation for the Stokes
vector is also derived and solved directly. In Sec. 5, the
Poisson-bracket formalism for BS is developed and shown to be
consistent with the Stokes-vector equation. In Sec. 6, the Pauli
spin-matrix formalism is used to relate the (unknown) elements of a
Jones-space transfer matrix to the (known) elements of a
Stokes-space rotation matrix. The mathematics of the bracket and
spin-matrix formalisms are similar to the mathematics of
angular-momentum operators. Hence, familiarity with the former
facilitates the transition from the classical model of BS to the
quantal model (which will be described elsewhere). Finally, in Sec.
7 the main results of this report are summarized.

The evolution of a monochromatic wave with two polarization
components was described, in Jones space and Stokes space, in
\cite{gor00}. The notation and results of \cite{gor00} will be used
without further comment.


\sec{2. Coupled modes}

The initial evolution of BS (during which the pumps are not
depleted) is governed by the Hamiltonian
\be H = \delta(|A_s|^2 - |A_i|^2) + \gamma A_s^*A_i +
\gamma^*A_sA_i^*, \label{2.1} \ee
together with the Hamilton equations
\be d_zA_j = i\partial H/\partial A_j^*. \label{2.2} \ee
For the case in which the pump and sideband polarizations are
parallel, the wavenumber mismatch $\delta = (\beta_s + \beta_q -
\beta_i - \beta_p)/2 + \gamma_K(|A_p|^2 - |A_q|^2)/2$, where
$\beta_j = \beta(\omega_j)$ are wavenumbers, $\gamma_K$ is the Kerr
coefficient and $A_p$ and $A_q$ are pump amplitudes, and the
coupling coefficient $\gamma = 2\gamma_KA_pA_q^*$ \cite{mck02}.
Other polarization configurations are discussed in
\cite{mck04a,mck06b}. By combining Eqs. (\ref{2.1}) and (\ref{2.2}),
one obtains the (linear) coupled-mode equations
\ba d_zA_s &= &i\delta A_s + i\gamma A_i, \label{2.3} \\
d_zA_i &= &i\gamma^* A_s - i\delta A_i. \label{2.4} \ea
Similar equations govern power transfer in a directional coupler
\cite{kog90} and sum-frequency generation in a crystal \cite{boy92}.
Equivalent equations for the sideband powers $P_j = |A_j|^2$ and
phases $\phi_j = \arg(A_j)$ are derived in the Appendix.

The solutions of Eqs. (\ref{2.3}) and (\ref{2.4}) can be written in
the input--output form
\ba A_s(z) &= &\mu(z)A_s(0) + \nu(z)A_i(0), \label{2.5} \\
A_i(z) &= &-\nu^*(z)A_s(0) + \mu^*(z)A_i(0), \label{2.6} \ea
where the transfer functions
\ba \mu(z) &= &\cos(kz) + i\delta\sin(kz)/k, \label{2.7} \\
\nu(z) &= &i\gamma\sin(kz)/k \label{2.8} \ea
and the BS wavenumber $k = (\delta^2 + |\gamma|^2)^{1/2}$. Notice
that the transfer functions satisfy the auxiliary equation $|\mu|^2
+ |\nu|^2 = 1$, which is a manifestation of power conservation. The
signal-to-idler conversion efficiency $|A_i(z)|^2/|A_s(0)|^2$
attains its maximum of $|\gamma|^2/k^2$ when $kz = \pi/2$.


\sec{3. Jones vector}

The scalar equations (\ref{2.3}) and (\ref{2.4}) can be rewritten as
the matrix equation
\be d_zA = iKA, \label{3.1} \ee
where the amplitude (Jones) vector $A = [A_s,A_i]^t$ and the
coefficient matrix
\be K = \left[\begin{array}{cc} \delta & \gamma \\
\gamma^* & -\delta \end{array}\right]. \label{3.2} \ee
Equation (\ref{3.1}) has the formal solution
\be A(z) = U(z)A(0), \label{3.3} \ee
where the transfer matrix $U = e^{iKz}$. Since $K$ is hermitian, $U$
is unitary. Hence, it can be written in the Caley--Klein form
\be U = \left[\begin{array}{cc} \mu & \nu \\
-\nu^* & \mu^* \end{array}\right], \label{3.4} \ee
which is consistent with Eqs. (\ref{2.5}) and (\ref{2.6}). This
result is a manifestation of power conservation. Notice that Eq.
(\ref{3.1}) is equivalent to the Hamiltonian
\be H = A^\d KA, \label{3.5} \ee
together with the vector Hamilton equation
\be d_zA = i\partial H/\partial A^\d. \label{3.6} \ee
%


\sec{4. Stokes vector}

Define $A_j = P_j^{1/2}e^{i\phi_j}$, where $P_j$ is a sideband power
and $\phi_j$ is a sideband phase. Then an inspection of Eq.
(\ref{2.1}) shows that the Hamiltonian depends on the phase
difference $\phi_s - \phi_i$, but not on the total phase $\phi_s +
\phi_i$. Hence, BS can be described in terms of three real variables
(not four). For such an interaction, it is natural to introduce the
Stokes-like parameters
\ba S_0 &= &|A_s|^2 + |A_i|^2, \label{4.1} \\
S_1 &= &|A_s|^2 - |A_i|^2, \label{4.2} \\
S_2 &= &A_sA_i^* + A_s^*A_i, \label{4.3} \\
S_3 &= &i(A_sA_i^* - A_s^*A_i), \label{4.4} \ea
in which the sideband amplitudes $A_s$ and $A_i$ play the roles of
the polarization components $X$ and $Y$ \cite{jac75}. $S_0$ is the
total sideband power, $S_1$ is the power difference, and $S_2$ and
$S_3$ contain information about the phase difference. Notice that
$S_0^2 = S_1^2 + S_2^2 + S_3^2$.

By combining Eqs. (\ref{2.5}) and (\ref{2.6}) with Eq. (\ref{4.1}),
one finds that
\be S_0(z) = S_0(0). \label{4.5} \ee
The total power is conserved because the Hamiltonian does not depend
on the total phase, as discussed in the Appendix. This result
implies that the Stokes vector $\vec{S} = (S_1,S_2,S_3)$ has
constant magnitude, so evolution in Stokes space is rotation on a
sphere of radius $S_0$ (the Stokes sphere). By combining Eqs.
(\ref{2.5}) and (\ref{2.6}) with Eqs. (\ref{4.2})--(\ref{4.4}), one
finds that
\be S(z) = R(z)S(0), \label{4.6} \ee
where the column vector $S = [S_1,S_2,S_3]^t$ and the rotation
matrix
\be R = \left[\begin{array}{ccc} (\mu_r^2 + \mu_i^2 - \nu_r^2 - \nu_i^2)
& 2(\mu_i\nu_i + \mu_r\nu_r) & 2(\mu_i\nu_r - \mu_r\nu_i) \\
2(\mu_i\nu_i - \mu_r\nu_r) & (\mu_r^2 - \mu_i^2 - \nu_r^2 + \nu_i^2)
& 2(\nu_r\nu_i + \mu_r\mu_i) \\
2(\mu_i\nu_r + \mu_r\nu_i) & 2(\nu_r\nu_i - \mu_r\mu_i) & (\mu_r^2 -
\mu_i^2 + \nu_r^2 - \nu_i^2)
\end{array}\right]. \label{4.7} \ee
The identity $|\mu|^2 + |\nu|^2 = 1$ ensures that $R$ is orthogonal.
By combining Eqs. (\ref{2.7}), (\ref{2.8}) and (\ref{4.7}), one
obtains the explicit formulas
\ba k^2R_{11}(\theta) &= &\delta^2 + (\gamma_r^2 + \gamma_i^2)\cos\theta, \label{4.8} \\
k^2R_{12}(\theta) &= &\delta\gamma_r(1 - \cos\theta) - \gamma_ik\sin\theta, \label{4.9} \\
k^2R_{13}(\theta) &= &-\delta\gamma_i(1 - \cos\theta) - \gamma_rk\sin\theta, \label{4.10} \\
k^2R_{21}(\theta) &= &\delta\gamma_r(1 - \cos\theta) + \gamma_ik\sin\theta, \label{4.11} \\
k^2R_{22}(\theta) &= &\gamma_r^2 + (\gamma_i^2 + \delta^2)\cos\theta, \label{4.12} \\
k^2R_{23}(\theta) &= &-\gamma_r\gamma_i(1 - \cos\theta) + \delta k\sin\theta, \label{4.13} \\
k^2R_{31}(\theta) &= &-\delta\gamma_i(1 - \cos\theta) + \gamma_rk\sin\theta, \label{4.14} \\
k^2R_{32}(\theta) &= &-\gamma_r\gamma_i(1 - \cos\theta) - \delta k\sin\theta, \label{4.15} \\
k^2R_{33}(\theta) &= &\gamma_i^2 + (\delta^2 +
\gamma_r^2)\cos\theta, \label{4.16} \ea
where the distance parameter $\theta = 2kz$. The evolution of the
Stokes vector is illustrated in Fig. 2.
\begin{figure}[h]
\centerline{\includegraphics[width=2.4in]{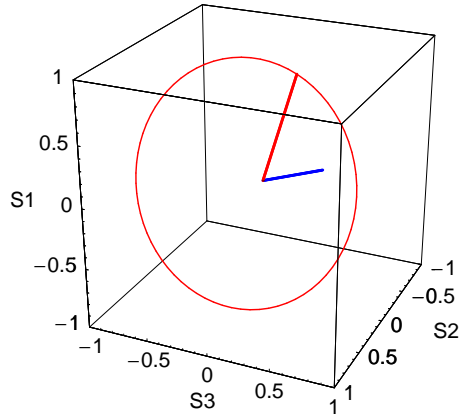}}
\vspace{-0.1in} \caption{Trajectory of the Stokes vector for
$\delta/\gamma_r = 0.3$ and $\gamma_i/\gamma_r = 0.1$. The thick
blue and red lines denote the rotation axis $\vec{r}$ and input
Stokes vector $\vec{S}(0)$, respectively. The thin red curve denotes
the trajectory of the tip of the Stokes vector.}
\end{figure}

In the preceding analysis, known results for the mode amplitudes
were used to deduce results for the Stokes parameters. However, one
can also determine the evolution of the Stokes parameters directly.
By combining Eqs. (\ref{2.3}) and (\ref{2.4}), one obtains the
evolution equations
\ba d_zS_0 &= &0, \label{4.17} \\
d_zS_1 &= &-2\gamma_iS_2 - 2\gamma_rS_3, \label{4.18} \\
d_zS_2 &= &2\delta S_3 + 2\gamma_iS_1, \label{4.19} \\
d_zS_3 &= &2\gamma_rS_1 - 2\delta S_2. \label{4.20} \ea
Equation (\ref{4.17}) is consistent with Eq. (\ref{4.5}), and Eqs.
(\ref{4.18})--(\ref{4.20}) can be rewritten as the vector equation
\be d_z\vec{S} = \vec{\Omega}\times\vec{S}, \label{4.21} \ee
where the rotation vector $\vec{\Omega} = 2(-\delta,
-\gamma_r,\gamma_i)$. [If one were to replace $\gamma$ by
$\gamma^*$, the rotation vector would be
$-2(\delta,\gamma_r,\gamma_i)$ and some formulas would be simpler.
However, the appearance of $\gamma A_i$ in the equation for $A_s$ is
standard.] Define $\vec{\Omega} = 2k\vec{r}$, where the wavenumber
$k$ was defined after Eq. (\ref{2.8}) and $\vec{r} =
(-\delta,-\gamma_r,\gamma_i)/k$ is a unit vector parallel to
$\vec{\Omega}$. Then $\vec{S}_\parallel =
\vec{r}(\vec{r}\cdot\vec{S})$ is parallel to the rotation vector,
$\vec{S}_\perp = \vec{S} - \vec{r}(\vec{r}\cdot\vec{S})$ is
perpendicular to the rotation vector and lies in the plane defined
by $\vec{r}$ and $\vec{S}$, and the auxiliary vector $\vec{S}_\times
= \vec{r}\times\vec{S}$ is perpendicular to the rotation vector and
the $rs$-plane. If the Stokes vector rotates about $\vec{r}$ through
the angle $\theta = 2kz$, $\vec{S}_\parallel$ remains constant and
$\vec{S}_\perp$ becomes $\cos\theta\vec{S}_\perp +
\sin\theta\vec{S}_\times$. By combining these facts, one obtains the
rotation formula
\be \vec{S}(\theta) = [\cos\theta + (1 -
\cos\theta)\vec{r}\vec{r}\cdot + \sin\theta\vec{r}\times]\vec{S}(0).
\label{4.22} \ee
Equation (\ref{4.22}) is consistent with Eqs.
(\ref{4.8})--(\ref{4.16}). The quantity in square brackets is the
rotation operator, written in dyadic (rather than matrix) form.


\sec{5. Poisson-bracket formalism}

Another inspection of Eq. (\ref{2.1}) shows that the Hamiltonian
\be H = \delta S_1 + \gamma_rS_2 - \gamma_iS_3, \label{5.1} \ee
where $S_1$--$S_3$ were defined in Eqs. (\ref{4.2})--(\ref{4.4}).
This equation can be rewritten in the compact form $H =
-\vec{\Omega}\cdot\vec{S}/2$, where $\Omega$ was defined after Eq.
(\ref{4.21}). Since the Hamiltonian depends only on the Stokes
parameters, it is natural to formulate the interaction in Stokes
space. For any Stokes parameter $S_j$, the rate of change
\be d_zS_j = \sum_k \Biggl( {\partial S_j \over \partial A_k} {dA_k
\over dz} + {\partial S_j \over \partial A_k^*} {dA_k^* \over dz}
\Biggr). \label{5.2} \ee
By combining Eqs. (\ref{2.2}) and (\ref{5.2}), one obtains the
Hamilton equation
\be d_zS_j = i\{S_j,H\}, \label{5.3} \ee
where the Poisson bracket
\be \{P,Q\} = \sum_k \Biggl( {\partial P \over \partial A_k}
{\partial Q \over \partial A_k^*} - {\partial P \over \partial
A_k^*} {\partial Q \over \partial A_k} \Biggr). \label{5.4} \ee
Notice that $\{Q,P\} = -\{P,Q\}$. Since $H$ depends on $S_k$, to
determine the consequences of Eq. (\ref{5.3}) one must first
calculate the Poisson brackets $\{S_j,S_k\}$. The results are
\be \{S_j,S_k\} = \pm 2iS_l, \label{5.5} \ee
where the $+$ ($-$) sign applies if $j$, $k$ and $l$ are in positive
(negative) cyclic order. By combining Eqs. (\ref{5.3}) and
(\ref{5.5}), one obtains the evolution equations
\ba d_zS_1 = \Omega_2S_3 - \Omega_3S_2, \label{5.6} \\
d_zS_2 = \Omega_3S_1 - \Omega_1S_3, \label{5.7} \\
d_zS_3 = \Omega_1S_2 - \Omega_2S_1, \label{5.8} \ea
which are equivalent to Eq. (\ref{4.21}). $S_0$ is constant because
$\{S_0,S_j\} = 0$. The Poisson-bracket formulation of classical
mechanics is described in \cite{gol80}.


\sec{6. Spin-matrix formalism}

In the preceding sections the Stokes-space evolution was described
in terms of Jones-space transfer functions [Eqs. (\ref{2.7}) and
(\ref{2.8})], and directly in Stokes space [Eq. (\ref{4.21})]. In
this section the Jones-space evolution is described in terms of
Stokes-space quantities.

First, suppose that a Stokes vector is written in the polar form
\be \vec{S} = S_0(\cos\theta,\sin\theta\cos\phi,\sin\theta\sin\phi).
\label{6.1} \ee
Then the associated Jones vector
\be A =
S_0^{1/2}[\cos(\theta/2)e^{-i\phi/2},\sin(\theta/2)e^{i\phi/2}]^t.
\label{6.2} \ee
One can verify Eq. (\ref{6.2}) by combining it with Eqs.
(\ref{4.2})--(\ref{4.4}) and comparing the results to Eq.
(\ref{6.1}). Polar coordinates are illustrated in Fig. 3.
\begin{figure}[h]
\centerline{\includegraphics[width=2.4in]{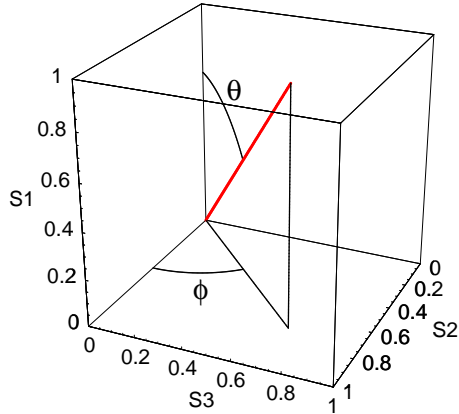}}
\vspace{-0.1in} \caption{Polar coordinates in Stokes space. The red
line denotes the Stokes vector $\vec{S}$.}
\end{figure}

Second, suppose that a Stokes-space rotation is specified by the
rotation axis $\vec{r}$ and angle $\theta$ (which should not be
confused with the polar angle of the preceding paragraph). Some
analysis is required to relate these quantities to their
counterparts in Jones space. In Sec. 3 it was shown that the
transfer matrix $U = e^{iKz}$, where $K$ is the coefficient matrix
(\ref{3.2}). One can facilitate the exponentiation of $iKz$ by
defining the identity matrix $\sigma_0$ and the Pauli spin matrices
\be \sigma_1 = \left[\begin{array}{cc} 1 & 0 \\ 0 & -1
\end{array}\right], \ \
\sigma_2 = \left[\begin{array}{cc} 0 & 1 \\
1 & 0 \end{array}\right], \ \
\sigma_3 = \left[\begin{array}{cc} 0 & -i \\
i & 0 \end{array}\right]. \label{6.3} \ee
These hermitian matrices have the properties $\sigma_j^2 = \sigma_0$
and $\sigma_j\sigma_k = \pm i\sigma_l$, where the $+$ ($-$) sign
applies if $j$, $k$ and $l$ are in positive (negative) cyclic order.
The latter result implies that $\sigma_j\sigma_k + \sigma_k\sigma_j
= 0$. Furthermore, any complex matrix $M$ can be written in the form
\be M = m_0\sigma_0 + \vec{m}\cdot\vec{\sigma}, \label{6.4} \ee
where the scalar and vector coefficients $m_0 = \tr(\sigma_0M)/2$
and $\vec{m} = \tr(\vec{\sigma}M)/2$, respectively, and the spin
vector $\vec{\sigma} = (\sigma_1,\sigma_2,\sigma_3)$. The
coefficient matrix has the decomposition
\be K = \delta\sigma_1 + \gamma_r\sigma_2 - \gamma_i\sigma_3,
\label{6.5} \ee
which can be rewritten in the compact form $K =
-k\vec{r}\cdot\vec{\sigma}$, where $k$ and $\vec{r}$ were defined
after Eq. (\ref{4.21}). By using the aforementioned properties of
the spin matrices, one finds that
\be U(z) = \sigma_0\cos(kz) - i\vec{r}\cdot\vec{\sigma}\sin(kz).
\label{6.6} \ee
By combining Eqs. (\ref{3.4}) and (\ref{6.4}), one also finds that
\be U = \mu_r\sigma_0 + i(\mu_i\sigma_1 + \nu_i\sigma_2 +
\nu_r\sigma_3). \label{6.7} \ee
It follows from Eqs. (\ref{6.5}) and (\ref{6.6}) that
\be \mu_r(z) = \cos(kz), \ \ [\mu_i(z),\nu_i(z),\nu_r(z)] =
-\sin(kz)\vec{r}, \label{6.8} \ee
where $[\mu_i,\nu_i,\nu_r]$ is a row vector. Equations (\ref{6.8})
relate the Jones-space transfer functions to the Stokes-space
rotation axis $\vec{r}$ and the distance parameter $kz$. The
discussion between Eqs. (\ref{4.21}) and (\ref{4.22}) shows that $kz
= \theta/2$, where $\theta$ is the Stokes-space rotation angle.

The connections between the formulas that describe evolution in
Jones and Stokes space are not accidental. It follows from Eqs.
(\ref{4.2})--(\ref{4.4}) and (\ref{6.3}) that
\be S_j = A^\d\sigma_jA. \label{6.9} \ee
By combining  Eqs. (\ref{3.1}) and (\ref{6.9}), one finds that
\be d_zS_j = iA^\d[\sigma_j,K]A, \label{6.10} \ee
where the commutator
\be [P,Q] = PQ - QP. \label{6.11} \ee
The spin matrices satisfy the commutation relations
\be [\sigma_j,\sigma_k] = \pm 2i\sigma_l, \label{6.12} \ee
where the $+$ ($-$) sign applies if $j$, $k$ and $l$ are in positive
(negative) cyclic order. Equations (\ref{6.12}) are the analogs of
Eqs. (\ref{5.5}) and the identity
$[\vec{\sigma},\vec{r}\cdot\vec{\sigma}] =
2i\vec{r}\times\vec{\sigma}$ follows from them. By combining this
identity with Eq. (\ref{6.10}), one obtains the evolution equation
\be d_z\vec{S} = 2k\vec{r}\times\vec{S}, \label{6.13} \ee
which is equivalent to Eq. (\ref{4.21}). $S_0$ is constant because
$[\sigma_0,\sigma_j] = 0$. Other connections between Jones space and
Stokes space are described in \cite{gor00}.


\sec{7. Summary}

Optical frequency conversion by Bragg scattering (BS) in a fiber was
considered. The evolution of BS can be modeled using the signal and
idler (sideband) amplitudes [Eqs. (\ref{2.3}) and (\ref{2.4})],
which are complex, or the sideband powers and phases [Eqs.
(\ref{a3})--(\ref{a6})], which are real. The amplitudes are the
components of a Jones vector and their spatial evolution is a linear
transformation in Jones space [Eq. (\ref{3.3})]. However, the
Hamiltonian for BS [Eq. (\ref{2.1}) or (\ref{a1})] does not depend
on the total sideband phase. This fact has two important
consequences. First, BS can be described by only three real
variables, which are the components of a Stokes-like vector. The
Stokes-space description of BS is useful because it allows one to
visualize power and phase information simultaneously. Second, the
total sideband power is conserved. In Jones space the norm of the
Jones vector is constant, so the transformation is unitary [Eq.
(\ref{3.4})], whereas in Stokes space the length of the Stokes
vector is constant, so its evolution is rotation [Eq. (\ref{4.22})].

The analysis of BS proceeded in three phases. In the first phase,
the coupled equations for the sideband (mode) amplitudes were stated
[Eqs. (\ref{2.3}) and (\ref{2.4})] and solved [Eqs. (\ref{2.7}) and
(\ref{2.8})]. These scalar equations were rewritten as a matrix
equation for the Jones vector [Eq. (\ref{3.1})] and their solutions
were used to specify the associated transfer matrix [Eq.
(\ref{3.4})]. The Stokes parameters associated with the mode
amplitudes were defined and the (unknown) elements of a Stokes-space
rotation matrix were related to the (known) elements of a
Jones-space transfer matrix [Eq. (\ref{4.7})]. In the second phase,
an evolution equation for the Stokes vector was derived [Eq.
(\ref{4.21})], which confirmed that evolution in Stokes space is
rotation. This equation has a concise solution [Eq. (\ref{4.22})],
which relates the rotation axis and angle to the mismatch and
coupling coefficients, and the distance. The Poisson-bracket
formalism for BS was developed [Eq. (\ref{5.3})] and shown to be
consistent with the Stokes-vector equation [Eqs.
(\ref{5.6})--(\ref{5.8})]. In the third phase, the Pauli spin-matrix
formalism was used to solve the Jones-vector equation [Eq.
(\ref{6.6})]. This solution is consistent with the solutions of the
coupled-mode equations and relates the (unknown) elements of a
Jones-space transfer matrix to the (known) elements of a
Stokes-space rotation vector [Eq. (\ref{6.8})]. Thus, explicit
formulas were derived, which relate the Jones and Stokes pictures of
BS. The power-phase formulation of the BS equations is described in
the Appendix.

The mathematics of the bracket and spin-matrix formalisms are
similar to the mathematics of angular-momentum operators. Hence,
familiarity with the former facilitates the transition from the
classical model of BS to the quantal model (which will be described
in a future report).

\sec{Acknowledgment}

I thank H. Kogelnik for his constructive comments on the manuscript.

\newpage

\sec{Appendix: Phase plane}

Section 2 was based on the complex formulation of the Hamilton
function (\ref{2.1}) and equations (\ref{2.2}). This appendix is
based on the associated real formulation. Define $A_j =
P_j^{1/2}e^{i\phi_j}$ and $\gamma = |\gamma|e^{i\phi_\gamma}$. Then
the initial evolution of BS is governed by the Hamiltonian
\be H = \delta(P_s - P_i) + 2|\gamma|(P_sP_i)^{1/2}\cos(\phi_s -
\phi_i - \phi_\gamma), \label{a1} \ee
together with the Hamilton equations
\be d_zP_j = -\partial H/\partial\phi_j, \ \ d_z\phi_j = \partial
H/\partial P_j. \label{a2} \ee
The assumption that $\gamma$ is real is equivalent to the
assumptions that $\phi_s$ and $\phi_i$ are measured relative to
$\phi_\gamma/2$ and $-\phi_\gamma/2$, respectively. By combining
Eqs. (\ref{a1}) and (\ref{a2}), one obtains the power equations
\ba d_zP_s &= &2\gamma(P_sP_i)^{1/2}\sin(\phi_s - \phi_i), \label{a3} \\
d_zP_i &= &-2\gamma(P_sP_i)^{1/2}\sin(\phi_s - \phi_i) \label{a4}
\ea
and the phase equations
\ba d_z\phi_s &= &\delta + \gamma(P_i/P_s)^{1/2}\cos(\phi_s - \phi_i), \label{a5} \\
d_z\phi_i &= &-\delta + \gamma(P_s/P_i)^{1/2}\cos(\phi_s - \phi_i).
\label{a6} \ea
It follows from Eqs. (\ref{a3}) and (\ref{a4}) that the total power
$P_t = P_s + P_i$ is conserved (because $H$ depends on $\phi_s -
\phi_i$). It also follows from Eqs. (\ref{a3})--(\ref{a6}) that $H$
is constant (because it does not depend explicitly on $z$).

Define the power difference $p = (P_s - P_i)/P_t$, phase difference
$\phi = \phi_s - \phi_i$ and Hamiltonian $h = H/(\gamma P_t)$. In
the notation of Sec. 4, $p = S_1/S_0$ and $\phi$ is the angle
between the projection of $\vec{S}$ on the 23-plane and the 2-axis
(Fig. 3). Furthermore, let $\delta/\gamma \rightarrow \delta$ and
$2\gamma z \rightarrow z$. Then BS is governed by the normalized
Hamiltonian
\be h = \delta p + (1 - p^2)^{1/2}\cos\phi \label{a7}, \ee
together with the normalized Hamilton equations
\be d_z p = -\partial h/\partial\phi, \ \ d_z \phi =
\partial h/\partial p. \label{a8} \ee
By combining Eqs. (\ref{a7}) and (\ref{a8}), one obtains the
normalized power and phase equations
\ba d_z p &= &(1 - p^2)^{1/2}\sin\phi, \label{a9} \\
d_z \phi &= &\delta - p\cos\phi/(1 - p^2)^{1/2}, \label{a10} \ea
respectively. Equations (\ref{a9}) and (\ref{a10}) are consistent
with Eqs. (\ref{a3})--(\ref{a6}).

Phase diagrams associated with the Hamiltonian (\ref{a7}) are shown
in Fig. 4. The phase point $(p,\phi)$ moves in such a way that an
observer moving with it, and looking forward, keeps higher energies
on his left. For $\delta = 0.0$ the trajectories are librations. The
trajectory that starts at the point $(-\pi/2,1)$ is a straight line
to $(-\pi/2,-1)$, followed by a phase jump to $(\pi/2,-1)$, followed
by a straight line to ($\pi/2,1)$, followed by a jump back to
$(-\pi/2,1)$. For $\delta = 0.3$ some trajectories are librations,
whereas others are rotations. The trajectory that starts at
$(-\pi/2,1)$ is a curve to the point $(\pi/2,1)$, followed by a
phase jump back to $(-\pi/2,1)$, and the trajectory that starts at
$(-\pi/2,-1)$ is a curve to the point $(\pi/2,-1)$, followed by a
jump back to $(-\pi/2,-1)$.

\begin{figure}[h]
\centerline{\includegraphics[width=3.0in]{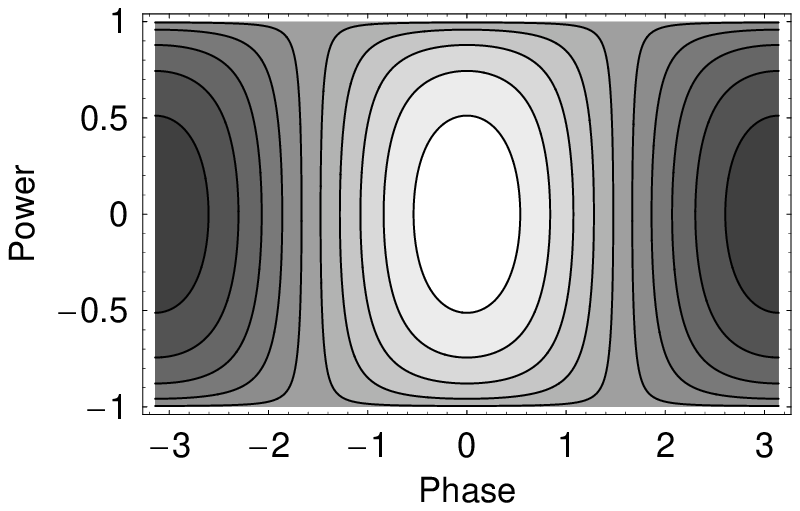}
\hspace{0.2in}
\includegraphics[width=3.0in]{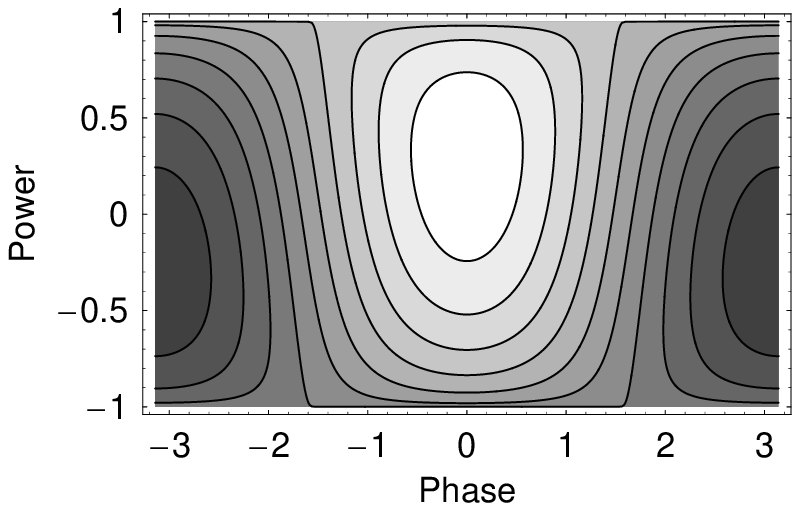}} \caption{Phase diagram for
($a$) $\delta/\gamma = 0.0$ and ($b$) $\delta/\gamma = 0.3$. Lighter
regions correspond to higher energies, whereas darker regions
correspond to lower energies. The energy contours (solid curves) are
the trajectories of the phase point $(p,\phi)$.}
\end{figure}

These phase diagrams describe the evolution of BS qualitatively. To
describe the evolution quantitatively, one must solve Eqs.
(\ref{a9}) and (\ref{a10}). First, suppose that $\delta = 0$, which
corresponds to (left-handed) rotation about the 2-axis in Stokes
space. Then
\ba (d_z p)^2 &= &(1 - p^2)\sin^2\phi, \nonumber \\
&= &1 - h^2 - p^2. \label{a11} \ea
It follows from Eq. (\ref{a11}) that
\be p(z) = (1 - h^2)^{1/2}\cos(z + z_0), \label{a12} \ee
where $h$ is constant and the distance parameter $z_0 =
\cos^{-1}[p(0)/(1 - h^2)^{1/2}]$. For the case in which $\delta = 0$
and $\gamma_i = 0$, Eqs. (\ref{4.6}), (\ref{4.8}) and (\ref{4.10})
can be rewritten in the normalized form
\be s_1(z) = s_1(0)\cos z - s_3(0)\sin z. \label{a13} \ee
By using the identity $s_3^2 + s_1^2 = 1 - s_2^2$, one finds that
$s_1(0) = (1 - s_2^2)^{1/2}\cos z_0$ and $s_3(0) = (1 -
s_2^2)^{1/2}\sin z_0$, where $s_2 = h$ and $z_0 =
\tan^{-1}[s_3(0)/s_1(0)]$ is the angle between the projection of
$\vec{S}(0)$ on the 31-plane and the 1-axis. Hence,
\be s_1(z) = (1 - s_2^2)^{1/2}\cos(z + z_0). \label{a14} \ee
Equation (\ref{a14}) is equivalent to Eq. (\ref{a12}).

Second, suppose that $\delta \neq 0$, which corresponds to
(left-handed) rotation about the axis $(\delta,1,0)/k$, where $k =
(1 + \delta^2)^{1/2}$. Then
\ba (d_z p)^2 &= &1 - p^2 - (h - \delta p)^2, \nonumber \\
&= &k^2[p_d^2 - (p - p_a)^2], \label{a15} \ea
where the power average $p_a = \delta h/k^2$ and power difference
$p_d = (k^2 - h^2)^{1/2}/k^2$. In the notation of Sec. 4, $p_a =
r_1(\vec{r}\cdot\vec{S})$ and $p_d = |\vec{S}_\perp(0)|/k$. It
follows from Eq. (\ref{a15}) that
\be p(z) = p_a + p_d\cos[k(z + z_0)], \label{a16} \ee
where $kz_0 = \cos^{-1}\{[p(0) - p_a]/p_d\}$. For the case in which
$\gamma_i = 0$, Eqs. (\ref{4.6}) and (\ref{4.8})--(\ref{4.10}) can
be rewritten in the normalized form
\ba k^2s_1(z) &= &s_1(0)[\delta^2 + \cos(kz)] + s_2(0)\delta[1 -
\cos(kz)] - s_3(0)k\sin(kz), \nonumber \\
&= &\delta[\delta s_1(0) + s_2(0)] + [s_1(0) - \delta
s_2(0)]\cos(kz) - ks_3(0)\sin(kz), \label{a17} \ea
where $\delta s_1(0) + s_2(0) = h$. By using the identity $s_1^2 +
s_2^2 + s_3^2 = 1$, one finds that  $[s_1(0) - \delta s_2(0)]^2 +
[ks_3(0)]^2 = k^2 - h^2$. Hence, Eq. (\ref{a17}) can be rewritten in
the form of Eq. (\ref{a16}), with $kz_0 =
\tan^{-1}\{ks_3(0)/[s_1(0)-\delta s_2(0)]\}$. The nonlinear
evolution of BS is described in \cite{mck93,ues02}.


\newpage

\begin{thebibliography}{99}

\bibitem{han02} J. Hansryd, P. A. Andrekson, M. Westlund, J. Li and P. O.
Hedekvist, ``Fiber-based optical parametric amplifiers and their
applications,'' IEEE J. Sel. Top. Quantum Electron. {\bf 8},
506--520 (2002).

\bibitem{mck07} C. J. McKinstrie, S. Radic and A. H. Gnauck, ``All-optical
signal processing by fiber-based parametric devices,'' Opt. Photon.
News {\bf 18} (3), 34--40 (2007).

\bibitem{lut92} G. G. Luther and C. J. McKinstrie, ``Transverse modulational
instability of counterpropagating light waves,'' J. Opt. Soc. Am. B
{\bf 9}, 1047--1060 (1992). Bragg reflection of counter-propagating
sidebands is discussed.

\bibitem{yu93} M. Yu, C. J. McKinstrie and G. P. Agrawal,
``Instability due to cross-phase modulation in the normal dispersion
regime,'' Phys. Rev. E {\bf 48}, 2178--2186 (1993). Bragg scattering
of co-propagating sidebands is mentioned.

\bibitem{mck93} C. J. McKinstrie, X. D. Cao and J. S. Li, ``Nonlinear detuning
of four-wave interactions,'' J. Opt. Soc. Am. B {\bf 10}, 1856--1869
(1993). If the four-wave equations are solved for three inputs with
arbitrary strengths, the solutions apply to both Bragg scattering
and phase conjugation.

\bibitem{ino94} K. Inoue, ``Tunable and selective wavelength conversion using
fiber four-wave mixing with two pump lights,'' IEEE Photon. Technol.
Lett. {\bf 6}, 1451--1453 (1994).

\bibitem{mar96} M. E. Marhic, Y. Park, F. S. Yang and L. G. Kazovsky, ``Widely
tunable spectrum translation and wavelength exchange by four-wave
mixing in optical fibers,'' Opt. Lett. {\bf 21}, 1906--1908 (1996).

\bibitem{mck02} C. J. McKinstrie, S. Radic and A. R. Chraplyvy, ``Parametric
amplifiers driven by two pump waves,'' IEEE J. Sel. Top. Quantum
Electron. {\bf 8}, 538--547 and 956 (2002).

\bibitem{ues02} K. Uesaka, K. K. Y. Wong, M. E. Marhic and L. G.
Kazovsky, ``Wavelength exchange in a highly nonlinear
dispersion-shifted fiber: Theory and experiments,'' IEEE J. Sel.
Top. Quantum Electron. {\bf 8}, 560--568 (2002).

\bibitem{tan04} T. Tanemura, C. S. Goh, K. Kikuchi and S. Y. Set,
``Highly efficient arbitrary wavelength conversion within entire
C-band based on nondegenerate fiber four-wave mixing,'' IEEE Photon.
Technol. Lett. {\bf 16}, 551--553 (2004).

\bibitem{mck04a} C. J. McKinstrie, H. Kogelnik, R. M. Jopson, S. Radic and
A. V. Kanaev, ``Four-wave mixing in fibers with random
birefringence,'' Opt. Express {\bf 12}, 2033--2055 (2004).




\bibitem{mck06b} C. J. McKinstrie, H. Kogelnik and L. Schenato,
``Four-wave mixing in a rapidly-spun fiber,'' Opt. Express {\bf 14},
8516--8534 (2006). This paper also reviews four-wave mixing in
strongly-birefringent and randomly-birefringent fibers.

\bibitem{gna06} A. H. Gnauck, R. M. Jopson, C. J. McKinstrie, J. C.
Centanni and S. Radic, ``Demonstration of low-noise frequency
conversion by Bragg scattering in a fiber,'' Opt. Express {\bf 14},
8989--8994 (2006).

\bibitem{mec06} D. M\'echin, R. Provo, J. D. Harvey and C. J. McKinstrie,
``180-nm wavelength conversion based on Bragg scattering in an
optical fiber,'' Opt. Express {\bf 14}, 8995--8999 (2006).

\bibitem{gor00} J. P. Gordon and H. Kogelnik, ``PMD fundamentals: Polarization
mode dispersion in optical fibers,'' Proc. Nat. Acad. Sci. {\bf 97},
4541--4550 (2000).

\bibitem{kog90} H. Kogelnik, ``Theory of optical waveguides,'' in {\it Guided-Wave
Optoelectronics, 2nd Ed.}, edited by T. Tamir (Springer, 1990),
Chapter 2.

\bibitem{boy92} R. W. Boyd, {\it Nonlinear Optics}\/ (Academic
Press, 1992), Chapter 2.

\bibitem{jac75} J. D. Jackson, {\it Classical Electrodynamics, 2nd.
Ed.} (Wiley, 1975), Chapter 7.

\bibitem{gol80} H. Goldstein, {\it Classical Mechanics, 2nd Ed.}
(Addison-Wesley, 1980), Chapter 9.

\end{thebibliography}
\end{document}